Complex diffusion mechanism of Zn in InP.

Rafał Jakiela

Institute of Physics, Polish Academy of Sciences, 02-668 Warszawa, Poland

Implantation and diffusion of Zn dopant into bulk InP:S were performed. Zn concentration profiles were investigated by secondary ion mass spectrometry and capacitance–voltage method. We find that thermodynamic conditions of annealing influence diffusion mechanism of Zn atoms. Excess phosphorus vacancies generation causes diffusion of Zn dopant by the complex $[V_P - Zn_{In} - V_P]$ producing characteristic diffusion profile with sharp slope. Such electrically inactive Zn configuration cause lower diffusion coefficient than that in generally known interstitial-substitutional mechanism.

* jakiela@ifpan.edu.pl (phone: +48 22 8437001 ex. 3434, fax: (+48 22) 843 09 26)

Introduction

Zinc is the most frequently used acceptor in InP based compounds. It is commonly accepted that zinc diffuses fastest by interstitial-substitutional mechanism with diffusion coefficient $D \sim C_S^{(m+1)}$ where $C_S$ is concentration of substitutional Zn and *m* is ionization state of zinc interstitials.

In this letter, we present results of closed ampoule Zn diffusion into S-doped InP. Annealing conditions allow observing two different diffusion mechanisms: generally observed interstitial-substitutional and never observed but earlier described [1] diffusion mechanism via complex $[V_P - Zn_{In} - V_P]$. Concentration profile for complex mechanism was compared with profile obtained from numerical simulation using the vacancy concentration dependent diffusion coefficient.

Experiment and results

Zn was diffused from implanted species. Annealing was performed in a sealed closed ampoule at the temperature of $700^O C$ for 15 minutes. Two thermodynamic conditions were applied: phosphorus overpressure or vacuum. Extent of diffusion of the species was characterized using the SIMS (Secondary Ion Mass Spectrometry) technique while the carriers concentration was evaluated by C-V profiling.

Figure 1 shows SIMS profiles from as-implanted sample and the samples annealed under different thermodynamic conditions. Zn profile in the sample annealed under phosphorus pressure, exhibits a well-known profile for interstitial-substitutional diffusion mechanism. Whereas, the profile in sample annealed in vacuum exhibits different shape, with constant atom concentration and sharper slope. We claim that such shape of Zn diffusion profile is produced by complex $[V_P-Zn_{In}-V_P]$ diffusion mechanism with diffusion coefficient dependent on phosphorus vacancy ($V_P$) concentration.

The dissociation reaction describing this mechanism can be written as

$$Zn_i^{+m} + V_{In} + 2V_P = [V_P - Zn_{In} - V_P] + mh \tag{1}$$

$Zn_i$ and $Zn_{In}$ indicates zinc interstitial and substitutional respectively, $V_{In}$ and $V_P$ indicates indium and phosphorus vacancy respectively, $h$ – hole concentration and superscript $m$ indicates charge state of zinc interstitial atoms. Using $c$ as a symbol of concentration, $k_1$ and $k_2$ for the reaction rate constants and $p$ for holes concentration, the law of mass action for reaction (1) is

$$c_i \cdot c_{VIn} \cdot c_{VP}^2 = k_1 \cdot c_n \cdot p^m \tag{2}$$

and for interstitial-substitutional mechanism [2]

$$c_i \cdot c_V = k_2 \cdot c_s \cdot p^{(m+1)} \tag{3}$$

Assuming one-dimensional diffusion model, diffusion equation for Zn atoms in particular configurations is

$$\frac{\partial c}{\partial t} = \frac{\partial}{\partial x}\left( D_s \frac{\partial c_s}{\partial x} + D_n \frac{\partial c_n}{\partial x} + D_i \frac{\partial c_i}{\partial x} \right) \tag{4}$$

where $D_s$, $D_n$, $D_i$ indicates diffusion coefficient for substitutional zinc atoms ($Zn_{In}$), zinc atoms in complex [$V_P$–$Zn_{In}$–$V_P$] and zinc interstitial respectively.

With equation (2) and (3) this can be transformed to

$$\frac{\partial c}{\partial t} = \frac{\partial}{\partial x}\left( \left( D_s + D_n k_1 \cdot c_s c_{VP}^2 + D_i k_2 \frac{c_s^{m+1}}{c_{VIn}} \right) \frac{\partial c_s}{\partial x} \right) \tag{5}$$

The above equation shows that Zn diffusion coefficient is indium and phosphorus vacancy as well as zinc concentration dependent. It also indicates that diffusion coefficient for interstitial-substitutional mechanism is dependent on Zn substitutional ($Zn_{In}$) and indium vacancy concentrations whereas for complex diffusion mechanism – on Zn substitutional and phosphorus vacancy concentrations.

In order to obtain diffusion profile for complex diffusion, we wrote computer simulation program, taking into account vacancy dependant diffusion coefficient. The program was written on the base of FTCS (Forward Time Center Space) method described by equation:

$$\frac{C_j^{n+1} - C_j^n}{\Delta t} = \frac{D_{j+\frac{1}{2}}(C_{j+1}^n - C_j^n) - D_{j-\frac{1}{2}}(C_j^n - C_{j-1}^n)}{(\Delta x)^2} \qquad (6)$$

$C$ indicates concentration of diffusing atoms, $D$ – diffusion coefficient, $j$ is numbering concentration points, $n$ – is numbering iteration steps, whereas $\Delta t$ and $\Delta x$ are time and depth steps respectively. Simulation condition set on this method is

$$\Delta t \leq \min_{j}\left[\frac{(\Delta x)^2}{2D_{j+\frac{1}{2}}}\right] \qquad (7)$$

Based on equation (5), diffusion coefficient for complex diffusion is dependent on factor $k_1 \cdot c_s \cdot c_{VP}^2$. Concentration profile of vacancy $V_P$ diffusion into semiconductor is described by equation [3],

$$C_V = (C_S - C_I) * erfc\left(\frac{x}{2\sqrt{dt}}\right) + C_I \qquad (8)$$

where $C_S$ and $C_I$ indicate surface and intrinsic vacancy concentration in semiconductor respectively. Concentration profiles for complex diffusion in sample with constant phosphorus vacancy concentration (equilibrium conditions annealing) and vacancy generation (vacuum annealing) obtained from computer simulation are shown in figure 2. Result for simulation of diffusion with variable vacancy concentration, exhibits similar profile shape to this achieved from SIMS measurement.

In order to compare profiles from SIMS measurement and computer simulation, we used in calculation real values for both surface concentration and diffusion coefficient of phosphorus vacancies, at the temperature of 700$^O$C. Surface vacancy concentration was calculated from Boltzmann equation

$$C_S = N \exp\left(\frac{-E_F}{kT}\right) \qquad (9)$$

on the base of energy formation data from [4] and was equal $4.7 \times 10^{18}$ cm$^{-3}$. Diffusion coefficient for phosphorus vacancy was $10^{-12}$ cm$^2$/s. Achieved simulation profile is with very good agreement with Zn profile obtained by SIMS method (Fig. 3)

Like interstitial-substitutional Zn diffusion in InP is forced by indium vacancy generation such complex [$V_P$–$Zn_{In}$–$V_P$] diffusion is forced by phosphorus vacancy generation. Described diffusion mechanism is also confirmed by C-V measurement. The majority of Zn atoms occupy then electrically inactive complex position and produce low concentration of acceptors in semiconductor (Fig. 4).

Conclusion

In conclusion, we have studied close-ampoule Zn diffusion in InP. Diffusion under phosphorus pressure (indium vacancy generation) is consistent with an interstitial-substitutional diffusion mechanism. However, annealing in a vacuum (phosphorus vacancy generation) cause switch to complex phosphorus vacancy concentration diffusion mechanism with lower diffusion coefficient. A tentative model based on diffusion equation and mass action law is proposed to explain complex diffusion mechanism.

**Figures**

Fig. 1 Zn concentration profiles for both as implanted and samples annealed under different condition at the temperature of 700$^O$C.

Fig. 2 Concentration profiles obtained from computer simulation for constant and variable phosphorus vacancy concentration.

Fig. 3 Zn concentration profiles for complex diffusion obtained from SIMS measurement and computer simulation.

Fig. 4 Zn atoms and p-type carriers concentrations profiles in sample annealed in vacuum.

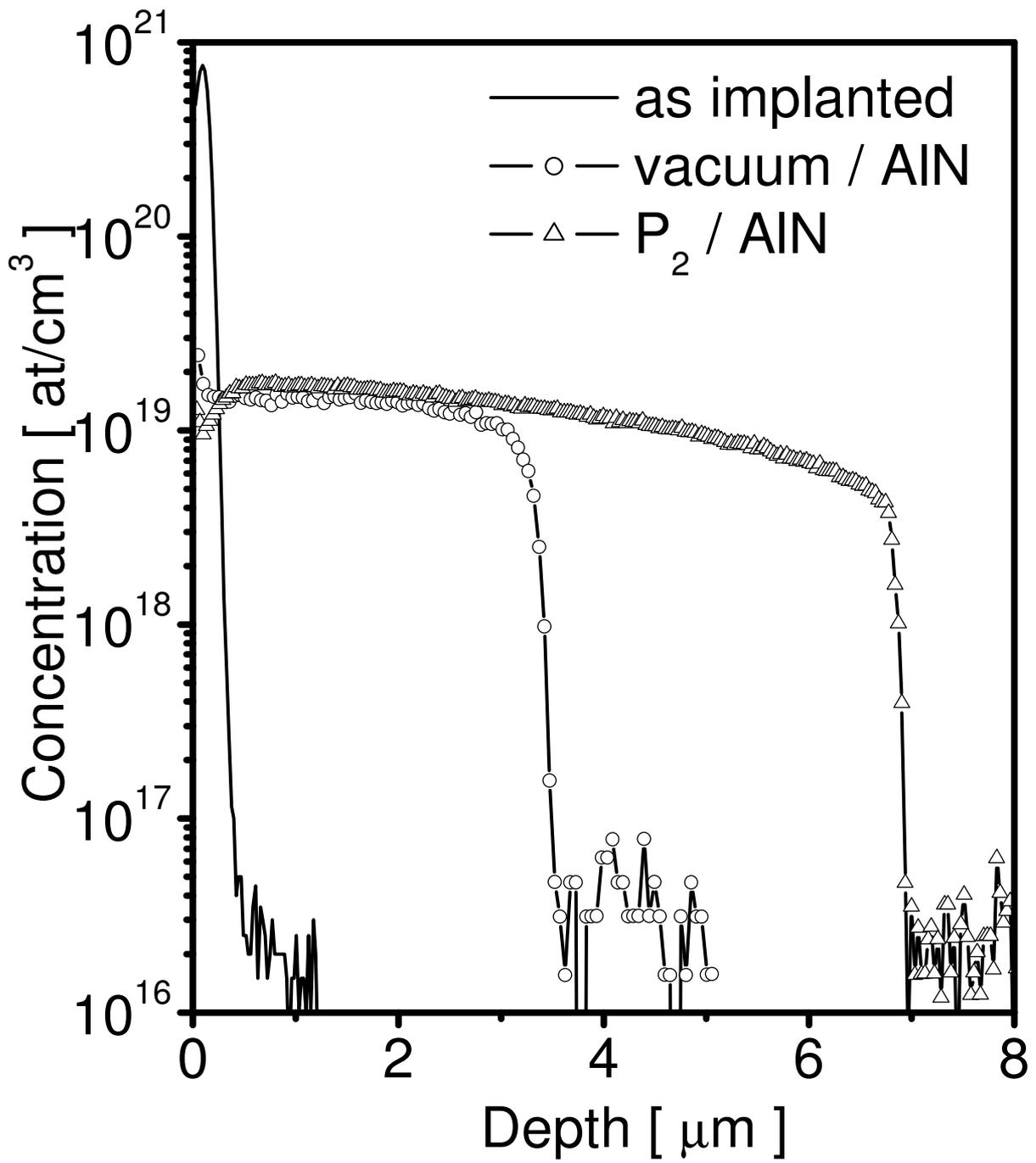

Fig. 1 Zn concentration profiles for both as implanted and samples annealed under different condition at the temperature of 700°C.

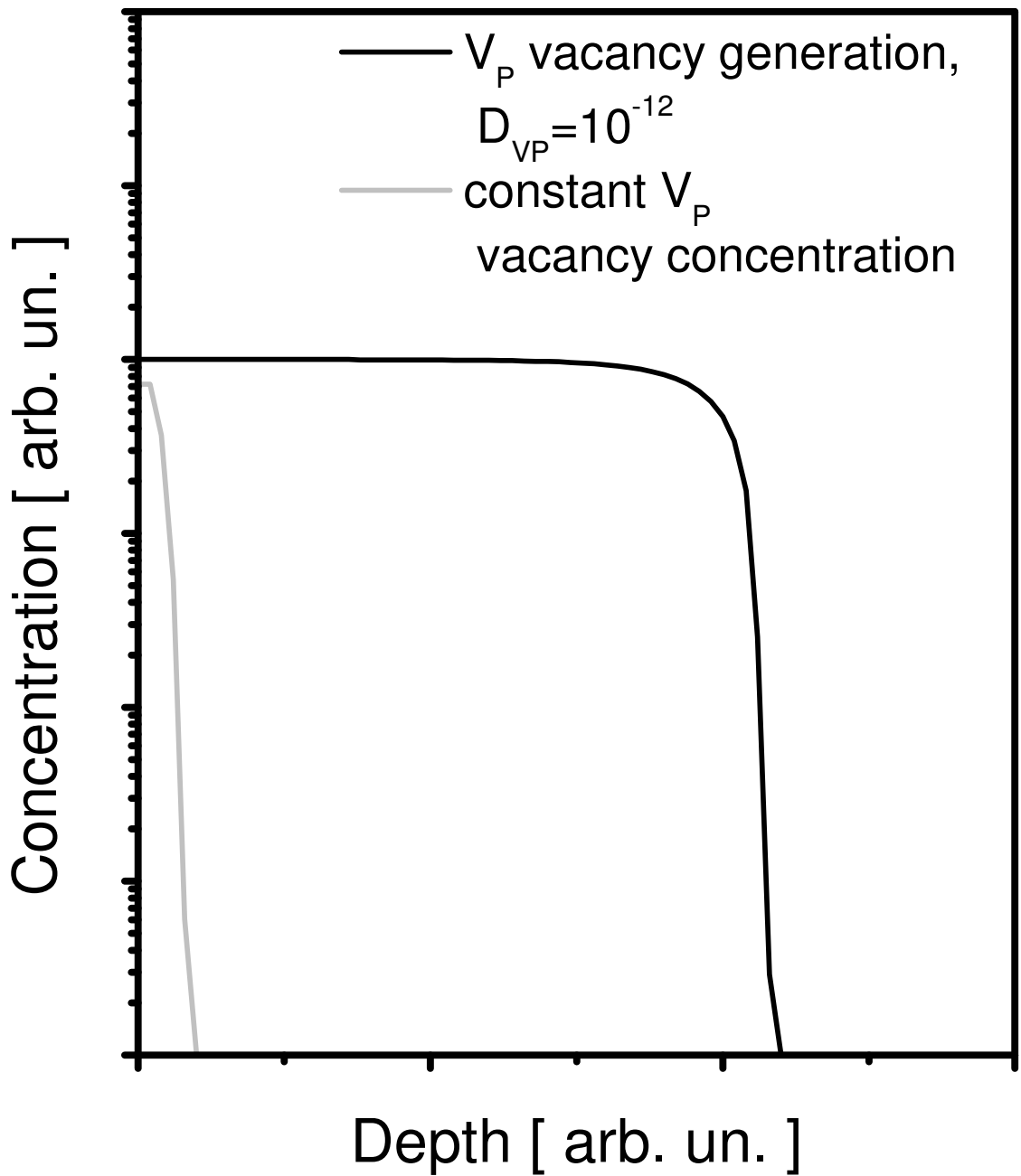

Fig. 2 Concentration profiles obtained from computer simulation for constant and variable phosphorus vacancy concentration.

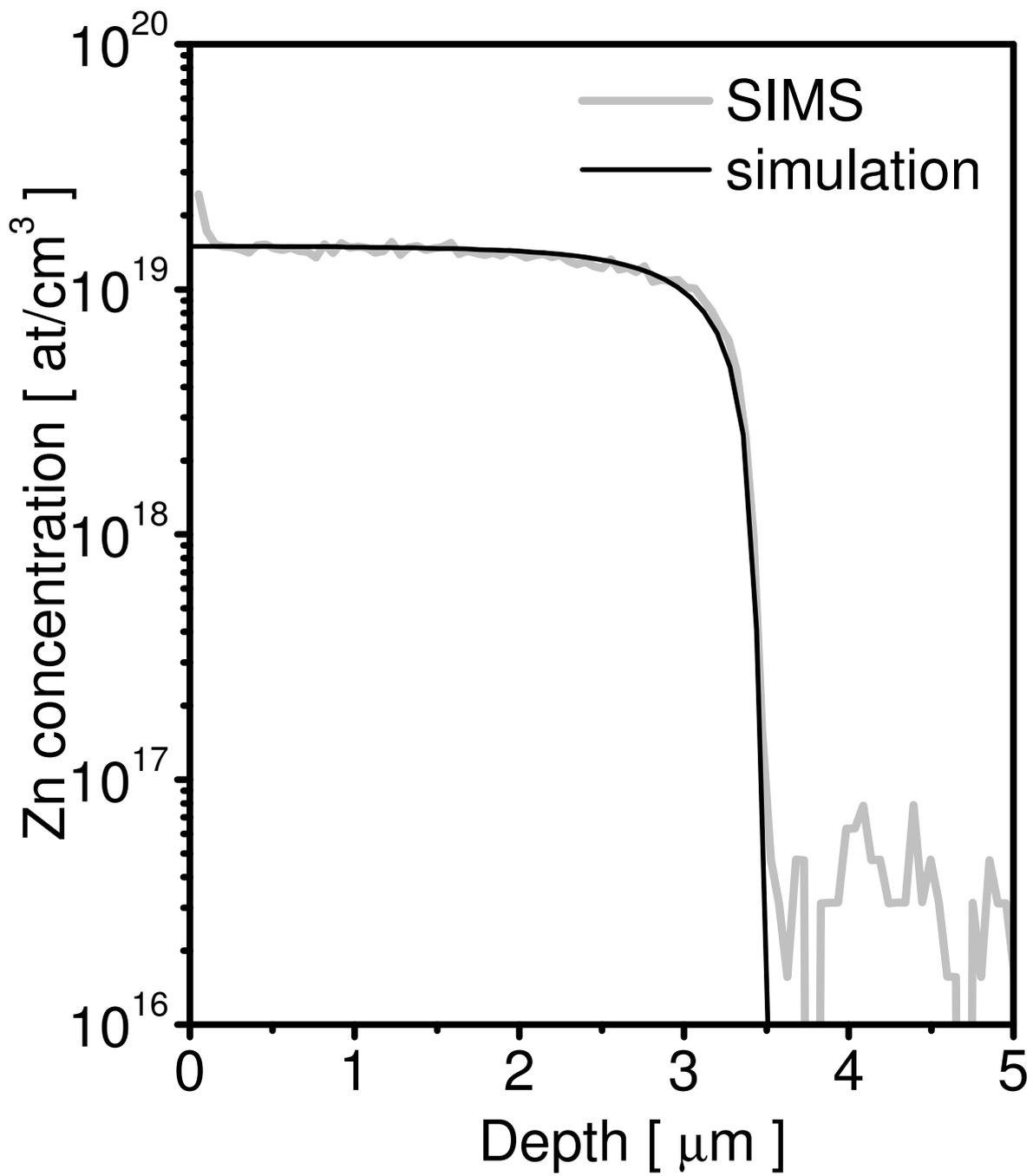

Fig. 3 Zn concentration profiles for complex diffusion obtained from SIMS measurement and computer simulation.

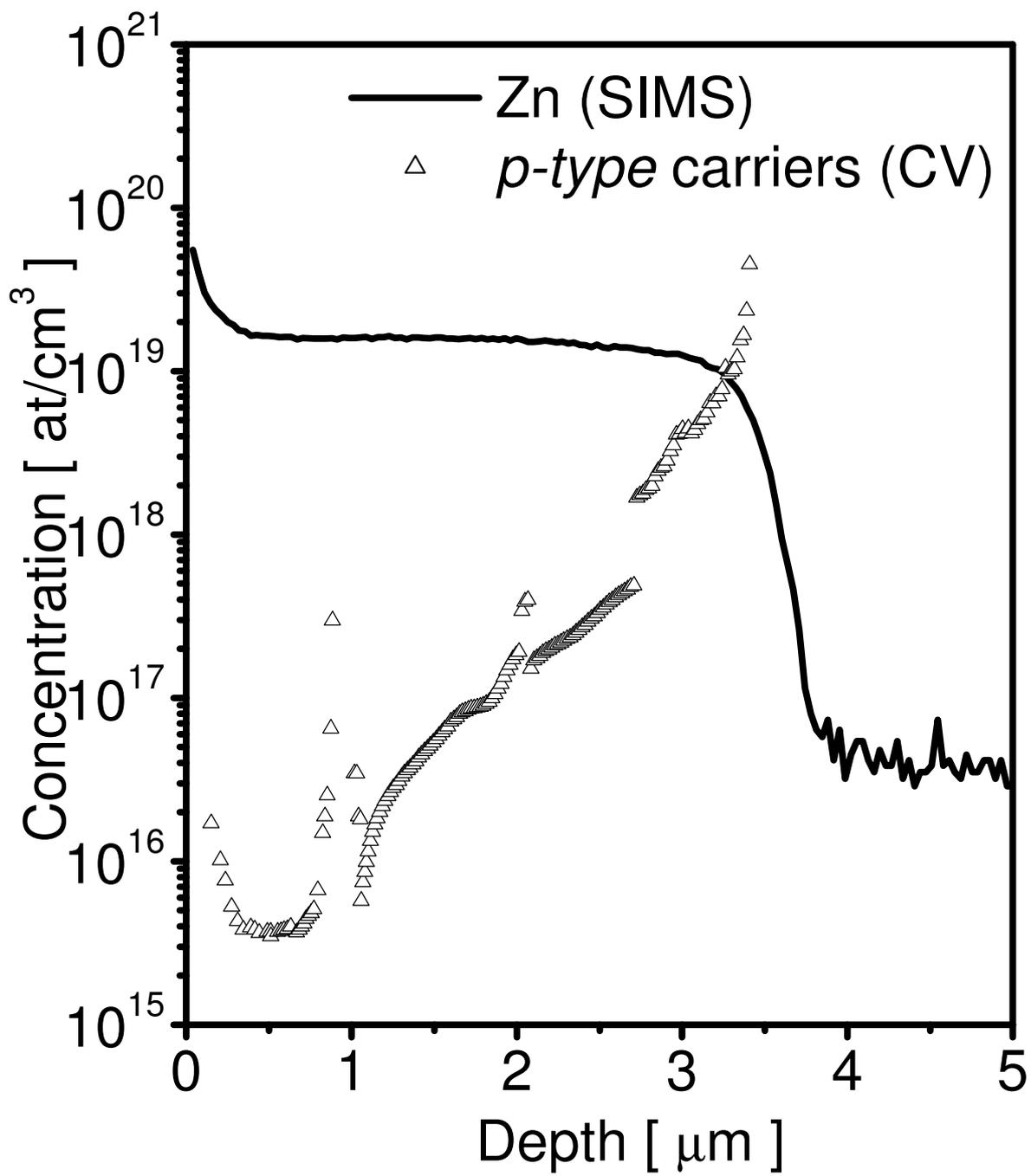

Fig 4. Zn atoms and p-type carriers concentrations profiles in sample annealed in vacuum.